# An Ultraluminous X-ray Source Powered by An Accreting Neutron Star

M. Bachetti[1,2], F. A. Harrison[3], D. J. Walton[3], B. W. Grefenstette[3], D. Chakrabarty[4], F. Fürst[3], D. Barret[1,2], A. Beloborodov[5], S. E. Boggs[6], F. E. Christensen[7], W. W. Craig[8], A. C. Fabian[9], C. J. Hailey[10], A. Hornschemeier[11], V. Kaspi[12], S.R. Kulkarni[3], T. Maccarone[13], J. M. Miller[14], V. Rana[3], D. Stern[15], S. P. Tendulkar[3], J. Tomsick[6], N. A. Webb[1,2], W. W. Zhang[11]

[1]Université de Toulouse, UPS-OMP, IRAP, F-31400 Toulouse, France

[2]CNRS; Institut de Recherche en Astrophysique et Planétologie; 9, Avenue du Colonel Roche, BP 44346, 31028 Toulouse Cedex 4, France Toulouse, France.

[3]Cahill Center for Astrophysics, 1216 East California Boulevard, California Institute of Technology, Pasadena, California 91125, USA.

[4]MIT Kavli Institute for Astrophysics and Space Research, Massachusetts Institute of Technology, Cambridge, MA 02139, USA

[5]Physics Department, Columbia University. 538 W 120th Street, New York, NY 10027, USA

[6]Space Sciences Laboratory, University of California, Berkeley, CA 94720, USA

[7]DTU Space, National Space Institute, Technical University of Denmark, Elektrovej 327, DK-2800 Lyngby, Denmark

[8]Lawrence Livermore National Laboratory, Livermore, CA 94550, USA

[9]Institute of Astronomy, University of Cambridge, Madingley Road, Cambridge CB3 0HA, UK

[10]Columbia Astrophysics Laboratory, Columbia University, New York, NY 10027, USA

[11]NASA Goddard Space Flight Center, Greenbelt, MD 20771, USA

[12]Department of Physics, McGill University, Montreal, Quebec, H3A 2T8, Canada

[13]Department of Physics, Texas Tech University, Lubbock, TX 79409, USA

[14]Department of Astronomy, University of Michigan, 500 Church Street, Ann Arbor, MI 48109-1042, USA

[15]Jet Propulsion Laboratory, California Institute of Technology, Pasadena, CA 91109, USA

**Ultraluminous X-ray sources (ULX) are off-nuclear point sources in nearby galaxies whose X-ray luminosity exceeds the theoretical maximum for spherical infall (the Eddington limit) onto stellar-mass black holes[1,2]. Their luminosity ranges from $10^{39}$ erg s$^{-1}$ $\leq L_X$ (0.5 – 10 keV) $\leq 10^{41}$ erg s$^{-1}$ [3]. Since higher masses imply less extreme ratios of the luminosity to the isotropic Eddington limit theoretical models have focused on black hole rather than neutron star systems[1,2]. The most challenging sources to explain are those at the luminous end ($L_X \geq 10^{40}$ erg s$^{-1}$), which require black hole masses $M_{BH} > 50$ $M_\odot$ and/or significant departures from the standard thin disk accretion that powers bright Galactic X-ray binaries. Here we report broadband X-ray observations of the nuclear region of the galaxy M82, which contains two bright ULXs. The observations reveal pulsations of average period 1.37 s with a 2.5-day sinusoidal modulation. The pulsations result from the rotation of a magnetized neutron star, and the modulation arises from its binary orbit. The pulsed flux alone corresponds to $L_X(3 – 30$ keV$) = 4.9 \times 10^{39}$ erg s$^{-1}$. The pulsating source is spatially coincident with a variable ULX[4] which can reach $L_X(0.3 – 10$ keV$) = 1.8 \times 10^{40}$ erg s$^{-1}$. This association implies a luminosity ~100 times the Eddington limit for a 1.4 solar mass object, or more than ten times brighter than any known accreting pulsar. This finding implies that neutron stars may not be rare in the ULX population, and it challenges physical models for the accretion of matter onto magnetized compact objects.**

The brightest accretion-powered X-ray pulsars, AO538-66[5], SMC X-1[6], and GRO J1744-28[7], are variable, with reported luminosities up to $L_X(2-20$ keV$) \sim 10^{39}$ erg s$^{-1}$, at the low end of the range that defines ULXs. Such luminosities, exceeding the Eddington limit for a neutron star by a factor of about six, can be understood[8] as resulting from accretion of material at moderately super-Eddington rates through a disk that couples to the neutron star's strong dipolar magnetic field (surface fields of B~$10^{12}$ Gauss). At the magnetospheric (Alfvén) radius, material is funnelled along the magnetic axis, radiation escapes from the column's side, and radiation pressure is ineffective at arresting mass transfer[9,10]. Explaining ULXs that have $L_X > 10^{39}$ erg s$^{-1}$ with a ~1 $M_\odot$ compact object using typical accreting pulsar models is however extremely challenging[8].

The NuSTAR (Nuclear Spectroscopic Telescope Array) high energy X-ray mission[11] observed the galaxy M82 (d≈3.6 Mpc) seven times between 2014 Jan 23 and 2014 Mar 06 as part of a follow-up campaign of the supernova SN2014J. The galaxy's disk contains several ULXs, the most luminous being M82X-1[12], which can reach $L_X$ (0.3-10 keV) ~$10^{41}$ erg s$^{-1}$, and the second brightest being a transient, M82X-2 (also referred to as X42.3+59[13]), which has been observed to reach[4,14] $L_X$ (0.3 -10 keV) ≈ 1.8 x $10^{40}$ erg s$^{-1}$. The two sources are separated by 5″, and so can only be clearly resolved by the Chandra X-ray telescope. During the M82 monitoring campaign, NuSTAR observed bright emission from the nuclear region containing the two ULXs. The region shows moderate flux variability at the 20% level during the first 22 days of observation. The flux then decreases by 60% during the final observation ~20 days later. The peak flux, $F_X$(3 – 30 keV) = (2.33 +/- 0.01) x $10^{-11}$ erg cm$^{-2}$ s$^{-1}$ (90% confidence; Figure 1) corresponds to a total 3 – 30 keV luminosity assuming isotropic emission of $3.7^{+0.01}_{-0.02} \times 10^{40}$ erg s$^{-1}$.

A timing analysis revealed a narrow peak just above the noise in a power density spectrum at a frequency of ~0.7 Hz. An accelerated epoch folding search[15] on overlapping 30 ksec intervals of data found coherent pulsations with a mean period of 1.37 s modulated with a sinusoidal period of 2.53 days throughout the 10-day interval starting at modified Julian day (MJD) 56691 (2014 Feb 03), and also in the last observation at MJD 56720 (Figure 1). The statistical significance of the pulsations is ~13-σ during the most significant 30 ksec segments, and >30-σ for the entire observation. A refined analysis (see Methods) subsequently enabled the detection of pulsations over a longer interval beginning on MJD 56686. The pulsed flux is variable, ranging from 5 – 13% (3 – 30 keV), and 8 – 23% (10 – 30 keV); see Figure 1 bottom panel. While the pulsed flux increases with energy this may result from a reduction in the contamination from other sources in the PSF rather than from a true increase in pulsed fraction. The maximum pulsed luminosity of the periodic source, NuSTAR J095551+6940.8, is $4.9^{+0.02}_{-0.03} \times 10^{39}$ erg s$^{-1}$ (3-30 keV).

Analysis of the period modulation yields a near-circular orbit (upper eccentricity limit of 0.003; see Methods) with a projected semi-major axis of 22.225(4) light-s (1-σ error). In addition to the orbital modulation, a linear spin-up of the pulsar is evident, with $\dot{P}$≈-2 x $10^{-10}$ s/s over the interval from MJD 56696 to 56701 when the pulse detection is most significant. Phase connecting the observations enables detection of a changing spin-up rate over a longer timespan (Figure 2 top panel) as well as erratic variations likely related to local changes in the torque on the neutron star applied by accreting matter[16].

Chandra observed M82 from MJD 56690.8396 to 56692.618, during an epoch when pulsations are detected. Only two sources in the Chandra image, M82X-1 and M82X-2 are sufficiently bright to be the counterpart of NuSTAR J095551+6940.8 (see Methods). We find the centroid of the pulsed emission ($\alpha = 09^h 55^m 51.05^s$, $\delta = +69°40'47.9''$) to be consistent with the location of M82X-2 (Figure 3). Monitoring by the *Swift* satellite establishes that the decrease in the nuclear region flux seen during ObsID011 is due to fading of M82X-1. The persistence of pulsations during this time further secures the association of the pulsating source, NuSTAR J095551+6940.8, with M82X-2. We derive a flux $F_X(0.5 - 10 \text{ keV}) = 4.07 \times 10^{-12}$ erg cm$^{-2}$s$^{-1}$, and an unabsorbed luminosity of $L_X(0.5 - 10 \text{ keV}) = (6.6 +/- 0.1) \times 10^{39}$ erg s$^{-1}$ for M82 X-2 during the Chandra observation.

The detection of coherent pulsations, a binary orbit, and spin-up behaviour indicative of an accretion torque unambiguously identify NuSTAR J095551+6940.8 as a magnetized neutron star accreting from a stellar companion. The highly circular orbit suggests the action of strong tidal torques, which, combined with the high luminosity, point to accretion via Roche lobe overflow. The orbital parameters give a mass function $f = 2.1$ M$_\odot$, and the lack of eclipses and assumption of a Roche-lobe-filling companion constrain the inclination to be i<60°. The corresponding minimum companion mass assuming a 1.4M$_\odot$ neutron star is M$_c$>5.2M$_\odot$, with radius R$_c$>7 R$_\odot$.

It is challenging to explain the high luminosity using standard models for accreting magnetic neutron stars. Adding the Chandra-measured E<10 keV luminosity to the E>10 keV pulsed flux (NuSTAR cannot directly spatially resolve the ULX), NuSTAR J095551+6940.8 has a luminosity $L_X(0.5 - 30 \text{ keV}) \sim 10^{40}$ erg s$^{-1}$. Theoretically, the X-ray luminosity depends strongly on the magnetic field and geometry of the accretion channel, being largest for a thin, hollow funnel that can result from the coupling of a disk onto the magnetic field[10]. A limiting luminosity $L_X \sim \frac{l_o}{2\pi d_o} L_{\text{Edd}}$, where l$_o$ is the arc length and d$_o$ the thickness of the funnel, can be reached if the magnetic field is high enough (B≥10$^{13}$ Gauss) to contain the accreting gas column[8]. Ratios of l$_o$/d$_o$~40 are plausible, so that the limiting luminosity can reach $L_X \sim 10^{39}$ erg s$^{-1}$, implying mass transfer rates exceeding Eddington by many times. Beyond this, additional factors increasing $L_X$ could result from increased $L_{Edd}$ due to very high (B>$10^{14}$ G) fields, which can reduce the electron scattering opacity[17], and/or a heavy neutron star. Some geometric beaming is also likely to be present.

This scenario is, however, difficult to reconcile with the measured rate of spin–up. The spin-up results from the torque applied by accreting material threading onto the magnetic field[18,19]. NuSTAR J095551+6940.8 is likely in spin equilibrium given the short spin-up timescale, $\frac{P}{\dot{P}} \sim 300\ yr$. Near equilibrium, the magnetosphere radius, $r_m$, is comparable to the co-rotation radius (the radius where a Keplerian orbit co-rotates with the neutron star), $r_{co} = \left(\frac{GM_{NS}P^2}{4\pi^2}\right)^{1/3} = 2.1 \times 10^8 \left(\frac{M_{NS}}{1.4 M_\odot}\right)\ cm$. With this assumption we can convert the measured torque, $\tau = 2\pi I\dot{\nu} = 6\times 10^{35}\ I_{45}\ g\,cm^2 s^{-2}$ (where $I_{45}$ is the neutron star moment of inertia $I/(10^{45}$ g cm$^2)$ and $\dot{\nu}$ is the measured frequency derivative) into a rate of matter magnetically channeled onto the pulsar, $\dot{M}_{mag} = 5\times 10^{-8} M_{1.4}^{-\frac{1}{2}} \left(\frac{r_m}{r_{co}}\right)^{-\frac{1}{2}} M_\odot yr^{-1} = 2.5 \left(\frac{r_m}{r_{co}}\right)^{-\frac{1}{2}} \dot{M}_{Edd}$. From the spin-up we therefore find an accretion rate that is only a few times higher than Eddington, independent of any assumption about the pulsar magnetic field. For an equilibrium spin period of 1.37 s and $\dot{M} \sim \dot{M}_{Edd}$, the implied magnetic field is B$\sim 10^{12}$ Gauss, typical of accreting pulsars in HMXBs, and too low to have an appreciable effect on $L_{Edd}$. It is possible that the current $\dot{M}$ is significantly larger than the average for this system, so that $r_m < r_{co}$, increasing $\dot{M}_{mag}$. For $\dot{M} = 100\ \dot{M}_{avg}$, $\dot{M}_{mag}$ is increased tenfold, so that only moderate geometric beaming is required to explain the observed luminosity. A fan beam geometry[9] viewed at a favorable angle could in this case produce the observed pulse profile (Figure 1) and provide the requisite moderate collimation.

The discovery of an ultra-luminous pulsar has implications for understanding the ULX population. The fraction of ULXs powered by neutron stars must now be considered highly uncertain. M82 X-2 has been extensively studied[4,14], however pulsations have eluded detection due to the limited timing capabilities of sensitive X-ray instruments, the transient nature of the pulsations, and the large amplitude of the orbital motion. Pulsars may indeed not be rare among the ULX population.

**Acknowledgements** This work was supported by NASA under grant no. NNG08FD60C, and made use of data from the Nuclear Spectroscopic Telescope Array (NuSTAR) mission, a project led by Caltech, managed by the Jet Propulsion Laboratory and funded by NASA. We thank the NuSTAR operations, software and calibration teams for support with execution and analysis of these observations. This work made use of data supplied by the UK Swift Science Data Centre at the University of Leicester. MB wishes to thank the Centre National d'Études Spatiales (CNES) and the Centre National de la Recherche Scientifique (CNRS) for the support.

**Author Contributions** M.B. reduction and timing analysis of the NuSTAR observations, interpretation of results, manuscript preparation. F.A.H.: interpretation of results, manuscript preparation. D.J.W.:

NuSTAR and Chandra spectroscopy, point source analysis. B.G.: NuSTAR image analysis. D.C.: accretion torque analysis, interpretation. F.F. verification of timing analysis, interpretation, D.B., A.B., A.C.F., A.H., V.M.K., T.M., J.T.: interpretation of results and manuscript review. S.B., F.C., W.W.C., C.J.H., D.S., S.P.T, N.W, W.W.Z.: manuscript review.

**Author Information** Reprints and permissions information is available at www.nature.com/reprints. The authors declare no competing financial interests. Readers are welcome to comment on the online version of the paper. Correspondence and requests for materials should be addressed to M.B. (matteo.bachetti@irap.omp.eu) and F.A.H. (fiona@srl.caltech.edu).

**FIGURES**

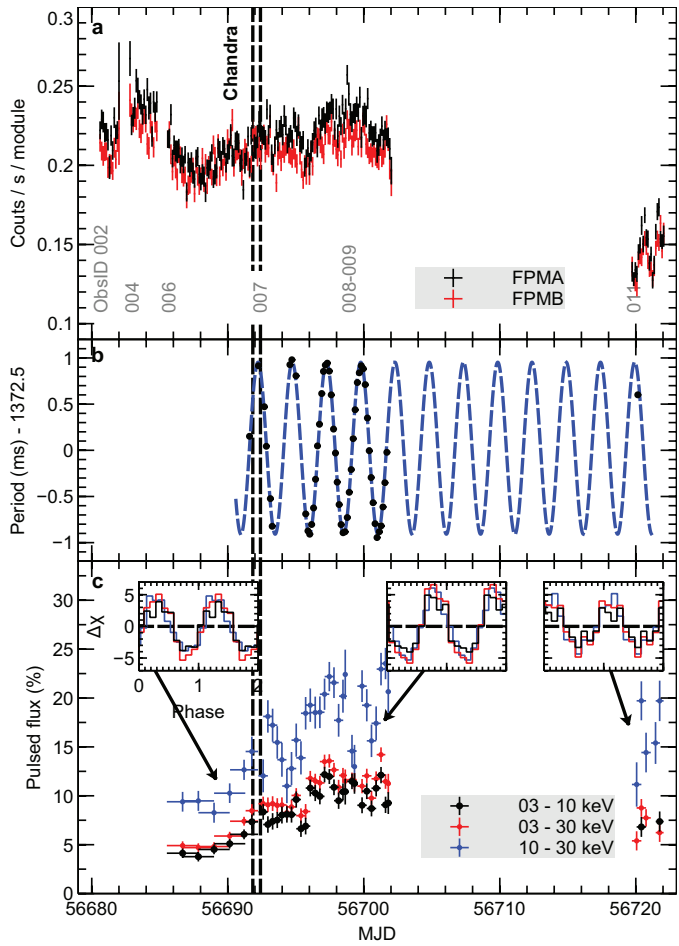

**Figure 1 The X-ray lightcurve and pulsations from the region containing NuSTAR J095551+6940.8.** Panel a: the background-subtracted 3 – 30 keV lightcurve extracted from a 70″-radius region around the position of NuSTAR J095551+6940.8. Black and red indicate the count rate from each of the two NuSTAR telescopes (1-σ errors). Panel b: detections of the pulse period (black points) fit using the best sinusoidal ephemeris (grey dashed line). The mean period is 1.37252266(12) seconds, with an orbital modulation period of 2.51784(6) days. The dashed vertical lines delineate the contemporaneous Chandra observation. Panel c shows the pulsed flux as a fraction of the emission from the 70″ region. The inserts show the pulse profile.

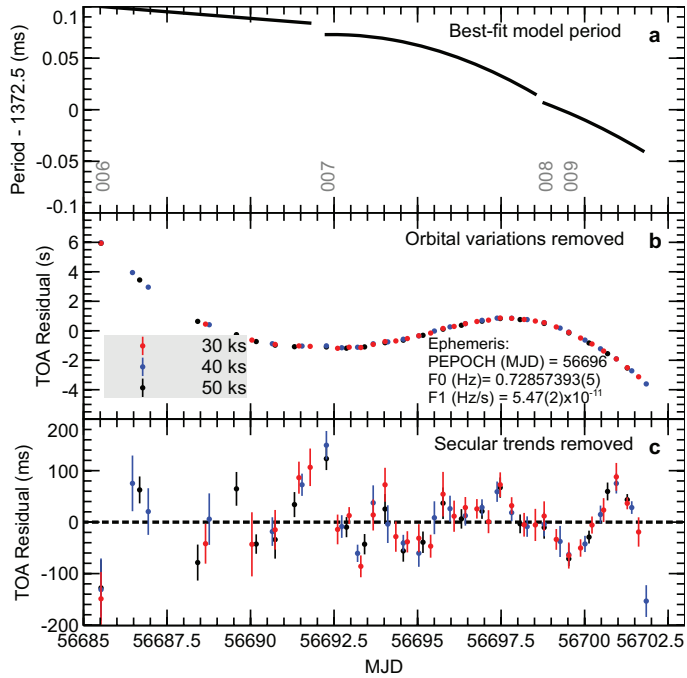

**Figure 2 The spin up behaviour of NuSTAR J095551+6940.8.** Panel a: the residual period after correcting for the sinusoidal orbital modulation given in Extended Data Table 2. The period decreases consistently, but the spin up rate is changing. Panel b: time of arrival residuals after correcting using the best-fit sinusoidal orbital modulation and constant period derivative. There is a clear trend independent of the choice of time binning (30, 40 or 50 ksec) that results from the variable spin up. Panel c: residuals after a smooth curve is fit to the TOA residuals. Residual noise remains in the TOAs at the 100 ms level (1-σ uncertainties).

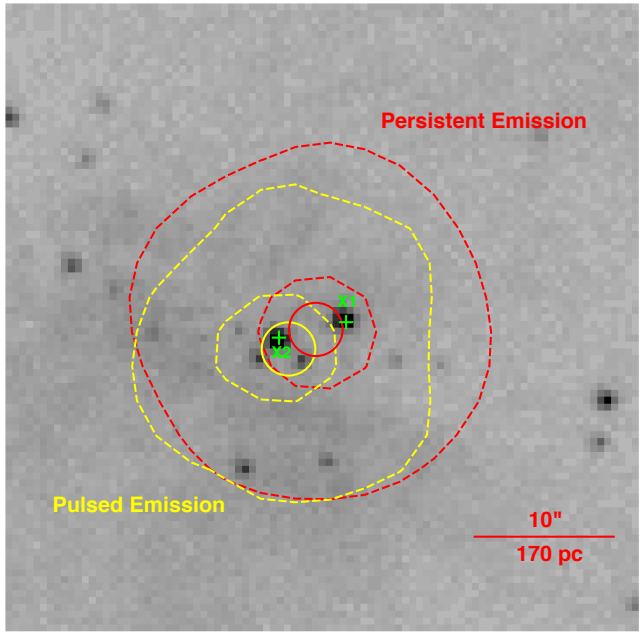

**Figure 3 The counterpart of NuSTAR J095551+6940.8.** The grayscale image shows a 45″ x 45″ Chandra image of the galaxy's center. Green diamonds mark the locations of M82 X-1 and X-2. NuSTAR 10-40 keV intensity contours (dashed) (50% and 90% levels) are shown for the pulsed (blue) and the persistent (red) emission. Solid error circles indicate the 3-σ statistical uncertainty on the centroid locations (see Methods). The pulsed emission centroid is consistent with the location of M82 X-2, and the centroid of the persistent emission is between M82 X-1 and X-2, indicating that there is additional persistent emission from X-2 as well as the persistent emission from X-1.

## METHODS

### Observations and preliminary data reduction

***NuSTAR***: *NuSTAR*[11] observed the M82 field 7 times between 2014 Jan 23 and 2014 Mar 06 (see **Extended Data Table 1** for details), for a total exposure of 1.91 Ms. We used the *NuSTAR* data analysis software (NuSTARDAS) version 1.2.0 and *NuSTAR* CALDB version 20130509 with the standard filters to obtain good time intervals, excluding the periods where the source was occulted by the Earth or was transiting through the South Atlantic Anomaly. *NuSTAR* records event arrival times with a resolution of 10 μs. Clock drifts, mostly due to temperature, are recorded at each ground station passage into a clock correction file that is updated monthly. The final time accuracy that can be reached by applying these clock corrections is ~2 ms. We applied these corrections to obtain solar system barycenter corrected event times using the general-purpose FTOOL barycorr and the NuSTAR clock correction file (v030).

***Chandra:*** *Chandra*[20] observed the M82 field using the ACIS-S detector in timed exposure (TE)/VFAINT mode between 2014-02-13 20:09:32 – 2014-02-04 14:50:00 for a total exposure of 47 ksec. We reduced the data with the standard pipeline in the *Chandra* Interactive Analysis of Observations software package (CIAO, version 4.6), filtering the data for periods of high background to produce a cleaned event list. We extracted point source spectra from circular regions of radius ~1-2″, depending on the proximity of other nearby sources, while the background was estimated from a larger region of radius ~35″, free of contaminating point sources and away from the plane of the M82 galaxy. Spectra and instrumental responses were produced from the cleaned events with the CIAO SPECEXTRACT tool, and were corrected for the fraction of the PSF falling outside the source region.

The primary goal of the *Chandra* observation was to constrain any faint soft X-ray emission from the recent SN2014J[21], and so the observation was performed with the maximum frame-time of 3.2s. Unfortunately, the two ULXs in M82 (X-1 and X-2) are sufficiently bright that the long frame integration time resulted in the *Chandra* spectra from these sources suffering from fairly severe pileup[22] (pileup refers to the scenario in which more than one photon is incident on a detector pixel within one frame time, resulting in the spurious apparent detection of a single photon with the combined energy of the individual incident photons). The detected counts-per-frame is ~0.45 for both X-1 and X-2, which equates to a pileup fraction of ~20-30% based on the *Chandra* ABC guide to pileup (http://cxc.harvard.edu/ciao/download/doc/pileup_abc.pdf). This unfortunately prevents a straightforward estimation of the fluxes of these two sources from the *Chandra* data alone. We will return to this issue below. None of the other point sources are bright enough to suffer significantly from these effects.

There is a known flux calibration discrepancy of ~10% between *NuSTAR* and *Chandra*. We account for this discrepancy, and quote fluxes based on the *NuSTAR* zeropoint.

**Swift:** *Swift* monitored the field of SN2014J regularly from 2014 Jan 23 through 2014 April. We reduced the data using the *Swift* mission automated data analysis server[23].

**Timing Analysis**

To determine the mean period and orbital parameters we divided the observations into overlapping 30 ks intervals and ran an accelerated epoch folding search[15] in each interval. To do this we used the software PRESTO[15], originally designed for radio observations of pulsars. To adapt the *NuSTAR* data to this software we modified the *makebininf* script by Abdo and Ray, originally written to use PRESTO with data from the *Fermi* satellite. To assess the quality of detection, accounting for possible difference in statistics between radio and X-ray data, we ran the same search with a random distribution of initial guesses for the period, far from the observed pulse period, and looked at the distribution of $\sigma$ for the "best detections". More than 99% of false detections were below a PRESTO value of $8\sigma$, which we use to define an acceptable detection. With this criterion we detected pulsations over ~10 days of observations. We cross-checked the detection with independent software using ISIS[24].

The observed pulse period exhibits sinusoidal variations resulting from the binary orbital motion. Assuming a circular orbit, the observed period $p_{obs}$ varies according to $p_{obs} = p_{em} (1 + X\Omega \sin(\Omega(t - T_{90})))$, where $p_{em}$ is the period in the system of reference co-moving with the pulsar, X is the projected semi-major axis of the orbit in light seconds, $\Omega = 2\pi/P_{orb}$ is the orbital angular velocity and $T_{90}$ is the time of longitude 90° (or mid-eclipse) assuming the ascending node is at longitude 0.

During ObsIDs 008 – 009, the interval over which we fit for the orbital parameters, we find an additional trend in the phases which requires a period derivative $\dot{P} \approx -2 \times 10^{-10}$ s/s. *NuSTAR*'s relative timing accuracy is currently known to ~2 ms over time intervals comparable to these observations. There is evidence for clock drifts of ~ 0.4 ms on the ~97 min orbital timescale. Taking this drift into account, the observed period could have a systematic error up to ~$10^{-7}$ s. This is orders of magnitude below the observed trend (~0.1ms) that we associate with a period derivative.

The ephemeris determined above is not sufficiently precise to align the pulses throughout any of the observations. In order to refine it, we calculated Times of Arrival (TOA) of the pulsations in each 30 ks interval and searched for an ephemeris that connects their phases[25]. To do this, we used the software Tempo2[26]. We found an orbital solution that aligns the TOAs to better than 36 ms (r.m.s.) from MJD 56696 to MJD 56701, the interval with the most significant detections.

We looked for signatures of eccentricity using the ELL1 model in Tempo2, adequate for low eccentricity orbits. We used TOAs calculated on different timescales (from 10 ks to 40 ks) in order to account for possible timescale-related effects. We measured values of the eccentricity that were always consistent with 0 within 2σ, with a maximum eccentricity of 0.002. We used this estimate to obtain a rough upper limit on the eccentricity of 0.003. The parameters of this orbital solution (determined for ObsIDs 008 – 009) can be found in **Extended Data Table 2**. Since the orbit is circular, we used the convention $T_0 = T_{asc}$ and so $T_{90} = T_{asc} + P_{orb} / 4$.

We found that the above ephemeris was not valid before MJD 56696, with the residuals from the best fit rapidly departing from 0. Therefore, for the rest of the ObsIds, we used PRESTO with the previously determined orbital solution, and searched the period-$\dot{P}$ plane for more precise solutions inside individual ObsIds. We determined values of period and period derivative that align the pulsation to better than 10% of the period inside single ObsIds. By using each of these newly determined ephemerides as a first approximation and propagating to the earlier observation, we were able to detect the pulsation on a much longer interval, including the whole of ObsId 006. These local solutions are summarized in **Extended Data Table 3**, and are the used for the phase-resolved analysis below.

We looked for any modulation of the phases at the orbital frequency (or harmonics) that might hint at poorly determined orbital parameters or yet-unmodeled orbital parameters (e.g. eccentricity, orbital period decay). We first subtracted the long-term trend of the phases with an ad-hoc function, as implemented in Tempo2. The residuals of this curve have an erratic behavior with quite large departures from a constant on timescales of several days. We calculated the Lomb-Scargle periodogram[27] of the residuals, looking for significant signatures at the orbital period or factors. We found no significant modulations of this kind. This timing noise is therefore not of orbital origin.

**Counterpart Identification**

*Chandra Imaging*: Extended Data Figure 1 shows an image of the central region of M82 from *Chandra* in the 0.3 – 10 keV band made from the observation coincident with *NuSTAR* ObsID 006. Given the pileup issues suffered by X-1 and X-2, we focus here on estimating fluxes from the rest of the point source population. Spectra are extracted for each source individually following the method outlined above. Given that these sources are mostly rather faint, we systematically rebin their spectra to have a minimum of 5 counts per bin, and minimize the Cash statistic[28] when analyzing these data. All spectral analysis is performed with XSPEC v12.8.1[29], and unless stated otherwise, quoted parameter uncertainties are the 90% confidence intervals for one parameter of interest.

We fit each source with two simple spectral models, the first an absorbed power law and the second an absorbed accretion disk (DISKBB[30] in XSPEC), focusing on the 1-10 keV bandpass to minimize the effects of the strong diffuse plasma emission from the M82 galaxy[31], which peaks at very low energies. When applying the former model, we limit the photon index to be positive, and for the latter model, we limit the disk temperatures to be less than 5 keV. These simple models provide good phenomenological descriptions of the data for the majority of the sources, the only exception being source 15, and we compute fluxes from the 3-10 keV bandpass for each (i.e. the common bandpass between *Chandra* and *NuSTAR*). For source 15, the *Chandra* data strongly require a second, soft X-ray component, and we additionally include a low temperature thermal plasma, modeled with the MEKAL code[32], assuming solar abundances), and compute 3-10 keV fluxes based on these models. These *Chandra* fluxes are compared to the total *NuSTAR* flux in the 3-10 keV bandpass of $(1.50 \pm 0.03) \times 10^{-11}$ erg cm$^{-2}$ s$^{-1}$, which we compute with a phenomenological combination of a Gaussian iron K-α emission line and a Comptonized continuum model (CompTT[33]) fit to the portion of the *NuSTAR* data simultaneous with the *Chandra* observation, extracted from the 70" region shown in Extended Data Figure 1.

Extended Data Table 4 provides the results of the power-law fits for the brightest five sources in this population. The disk-blackbody fluxes are typically within 30% of the powerlaw results (the latter generally being higher). Even taking the conservative approach and adopting the slightly higher powerlaw fluxes, none of these sources contribute more than 5% of the 3-10 keV *NuSTAR* flux. Given the observed 3-10 keV pulse fraction at the time of the *Chandra* observation, it is clear that only X-1 and X-2 are bright enough to be the origin of these pulsations.

*NuSTAR Astrometry*: We use the "persistent" emission to determine an absolute astrometric correction, which we then apply to the "pulse-on" images to determine the location of the pulsed emission. We assume that the persistent emission is composed equally of emission from M82 X-1 and X-2 since the contemporaneous Chandra imaging shows that these dominate the image and have comparable flux. To eliminate contamination from diffuse soft X-ray emission, we limit the image to the energy range from 10 to 40 keV. We find that we can align the *NuSTAR* image with the mean position of M82 X-1 and X-2 by applying a shift in the image plane of two sky pixels (which have a plate scale of 2.45 arc seconds / pixels), which is a reasonable shift given *NuSTAR*'s astrometric accuracy[11].

After astrometric alignment we use the "gcntrd.pro" routine from the IDL AstroLib to determine the position of the pulsed emission using data within 30 arc second of the peak brightness. This represents a net sample of 2,500 counts in the subtracted image.

To assess the statistical uncertainty on the source centroid we perform a Monte Carlo simulation that distributes 2,500 counts in a simulated image according to the on-axis *NuSTAR* PSF. For each Monte Carlo run we compute the centroid as above and compare the result with the (known) input source location. The X and Y offset distributions from the Monte Carlo have means of zero and 1-sigma widths of 0.235 pixels (0.577 arc seconds), corresponding to a 90% confidence interval of +/- 0.95 arc seconds in both X and Y.

The resulting RA/DEC centroid of the *NuSTAR* pulsed image is (RA: 09$^h$55$^m$51.05$^s$, DEC: +69°40′47.9″, J2000) with a combined statistical and systematic error of +/- 5″ in the projected image plane.

*Swift Imaging*:   We perform imaging analysis on *Swift* observations made during ObsIDs 007 and 009, when pulsations are detected and the flux from the region is high (see Figure 1). We use the *Swift*-XRT automated data processing pipeline[23], obtaining all observations from UT 2014 Feb 02 through 2014 Feb 11 and producing an image in the 5 to 10 keV band (Extended Data Figure 2, left). We then compare this to the *Swift* snapshot observations extracted from 2014/03/07 through 2014/03/11 that occurred just after the OBSID 011, when the flux from the *NuSTAR* extraction region containing M82 X-1 and X-2 had decreased by 40%, but pulsations were still detected. When the flux is high the *Swift* imaging shows significant contributions from both M82 X-1 and X- 1, but after ObsID 011 M82 X-2 is dominant in the *Swift* images. This is consistent with a significant decrease in flux of M82 X-1.  The continued detection of pulsations indicates that they originate from M82 X-2.

**Flux measurements and spectroscopy of M82 X-1 and M82 X-2**

We estimate the 3 – 10 keV flux from M82 X-2 during our observations from the *Chandra* observation. As noted previously, the *Chandra* data for X-2 are quite piled-up, which results in strong degeneracies in the flux inferred from these data alone. By including the *NuSTAR* data, which does not suffer from these effects, it is possible to break this degeneracy. Since the *NuSTAR* aperture is much larger than the aperture used for the *Chandra* data, it includes the emission from a variety of other point sources, including X-1 (for which the *Chandra* data is also piled-up, resulting in similar issues), and also much of the diffuse M82 plasma emission. The contribution of these additional sources of emission to the *NuSTAR* data need to be accounted for before the flux of X-2 can reliably be estimated from the combined dataset. Throughout all our analysis, we include a neutral absorption component fixed at the Galactic column in the direction of M82[34] ($N_H$ = 5.04 x 10$^{20}$ cm$^{-2}$) and additionally allow for variable neutral absorption intrinsic to M82.

*Diffuse Emission:*  We assess the contribution from the diffuse plasma emission, and extract an integrated *Chandra* spectrum from the 70" *NuSTAR* aperture with SPECEXTRACT, excluding all the point sources identified in Extended Data Figure 1. For the subsequent analysis, we follow [31]. [31]We leave a detailed analysis of the diffuse emission during this observation for future studies; for our purposes, it is only important that a phenomenological model with three MEKAL components provides a good description of the data. We find that this diffuse emission contributes (20 ± 1) % of the observed 3-10 keV *NuSTAR* flux.

*Other Point Sources:* We model the emission from the additional X-ray source population (i.e. excluding X-1 and X-2) for inclusion in the joint fitting. None of these sources are bright enough to significantly suffer from pileup. We first focus on the brightest two sources from this additional X-ray population, sources 15 and 18, and fit their full 0.3-10.0 keV *Chandra* spectra with a phenomenological model composed of a low-temperature MEKAL plasma (to account for the diffuse emission) and an accretion disk continuum.

For the rest of the fainter point source population, we combine their lower quality individual *Chandra* spectra into a single average spectrum, and simply model this to determine their contribution. We apply the same accretion disk continuum model used above. The total contribution from point sources other than X-1 and X-2 is found to be (17 ± 1) % of the observed 3-10 keV *NuSTAR* flux.

*M82 X-1 and X-2*: Having characterized all the other sources of emission in the *NuSTAR* aperture, we can now constrain the fluxes of X-1 and X-2 (given the pileup issues in the *Chandra* data, both sources need to be analyzed in this fashion simultaneously). We construct a 'contamination' model for the *NuSTAR* data obtained simultaneously with the *Chandra* observation, which is the sum of our characterizations of the diffuse emission and the other point sources, described above. The parameters of this contamination model are all set to the best-fit values obtained in our separate analysis of these various emission components, and are not allowed to vary.

To this model, we add two additional continuum components to represent X-1 and X-2, both absorbed. For the continuum, we use the CompTT model, as with different parameter combinations this model has the flexibility to provide both powerlaw-like and blackbody-like spectra, as required by the data. We then construct the same CompTT models for the *Chandra* data for X-1 and X-2 individually, and link the parameters for these components between the *NuSTAR* and the respective *Chandra* spectra. Finally, the *Chandra* models are modified by a pileup kernel[22], which accounts for the spectral distortions introduced by these effects. The influence of these effects, characterized by the grade migration parameter $\alpha$ [22], is not known a-priori, and is free to vary independently for X-1 and X-2. Although this pileup kernel can introduce strong degeneracies in the intrinsic flux inferred, this simultaneous modeling of the *NuSTAR* and the *Chandra* data, and the careful modelling of all the other sources of emission in the *NuSTAR* aperture, allows us to account for the pileup in the *Chandra* data, while also requiring that the total flux does not exceed that in the *NuSTAR* aperture, breaking this flux degeneracy.

With this approach, we find that the 3-10 keV fluxes of X-1 and X-2 are relatively similar during this *Chandra* observation, contributing $42^{+3}_{-4}$ and $21^{+2}_{-1}$ % of the 3-10 keV *NuSTAR* flux, corresponding to 3-10 keV luminosities of $(10 \pm 1) \times 10^{39}$ and $4.9^{+0.6}_{-0.3} \times 10^{39}$ erg s$^{-1}$ respectively (for a distance to M82 of 3.6 Mpc). Using the measured spectrum and absorption column, we find an unabsorbed 0.5 – 10 keV luminosity of $6.6 \times 10^{39}$ erg s$^{-1}$ for X-2.

## EXTENDED DATA TABLES

| ObsId | Date start (UT) | Date end (UT) | Effective Exposure (ks) |
|---|---|---|---|
| 80002092002 | 2014-01-23 12:58:00 | 2014-01-24 23:01:05 | 65.818 |
| 80002092004 | 2014-01-25 19:53:58 | 2014-01-27 19:16:03 | 89.965 |
| 80002092006 | 2014-01-28 12:31:46 | 2014-02-04 05:29:48 | 309.525 |
| 80002092007 | 2014-02-04 06:07:53 | 2014-02-10 18:14:28 | 306.240 |
| 80002092008 | 2014-02-10 18:51:40 | 2014-02-11 12:00:24 | 33.648 |
| 80002092009 | 2014-02-11 12:37:59 | 2014-02-13 23:45:23 | 114.893 |
| 80002092011 | 2014-03-03 17:07:14 | 2014-03-06 01:02:13 | 110.925 |

**Extended Data Table 1 List of *NuSTAR* observations used in this analysis**

Start and stop dates are given, as is the effective exposure of each observation. The exposure has been corrected for the period during which the source was occulted by the Earth, periods during which the instrument was not taking data and for the rate-dependent deadtime when the instrument was processing events and not sensitive to new incident photons.

| Parameter | Value |
|---|---|
| Orbital period $P_{orb}$ (d) | 2.53260(5) |
| Projected semi-major axis $X$ (ls) | 22.225(4) |
| Mid-eclipse time $T_{90}$ (MJD) | 56695.3659(1) |

**Extended Data Table 2 Best-fit orbital parameters**

Orbital parameters determined by fitting a sinusoidal orbital modulation to obsids 008-009. Error ranges for the parameters are given at the 1-sigma level.

| ObsId | MJD | Period (s) | Period derivative (s/s) | Period 2nd derivative (s/s$^2$) | TOA Scatter (ms) |
|---|---|---|---|---|---|
| 006 | 56685.5 | 1.3726001(4) | -3.0(1)×10$^{-11}$ | 0 | 62 |
| 007 | 56692.2 | 1.3725728(4) | 8(3)×10$^{-12}$ | -4.2(1)×10$^{-16}$ | 31 |
| 008-009 | 56698.7 | 1.3725076(4) | -1.38(7)×10$^{-10}$ | -3.3(5)×10$^{-16}$ | 9 |
| 011 | 56719.8 | 1.3722225(6) | -2.73(7)×10$^{-10}$ | 0 | 14 |

**Extended Data Table 3 Best fit period and period derivatives for individual *NuSTAR* observations**

Pulse period and period derivatives fit to Time of Arrivals (TOA) data for individual observations after removal of the orbital modulation. Errors are 1-sigma, and the parentheses indicate the error on the last digit of the reported value.

| Source | $N_H$ (10$^{22}$ cm$^{-2}$) | Γ | 3-10 keV flux contribution (%) |
|---|---|---|---|
| 15 | 50 $^{+10}_{-20}$ | 3 ± 1 | 4.4 ± 0.6 |
| 18 | 3 ± 1 | 0.4 ± 0.3 | 4.2 ± 0.6 |
| 9 | 6 ± 1 | 2.8 ± 0.4 | 1.3 ± 0.2 |
| 6 | 0.5 ± 0.3 | 1.6 ± 0.2 | 1.2 ± 0.2 |
| 20 | 2 $^{+4}_{-1}$ | < 1.0 | 1.0 $^{+0.2}_{-0.4}$ |

**Extended Data Table 4 3 - 10 keV flux contributions for bright sources near the nucleus of M82**

Fluxes derived from power-law fits to the five brightest sources after M82 X-1 and X-2. Errors are 90% confidence.

**EXTENDED DATA FIGURES**

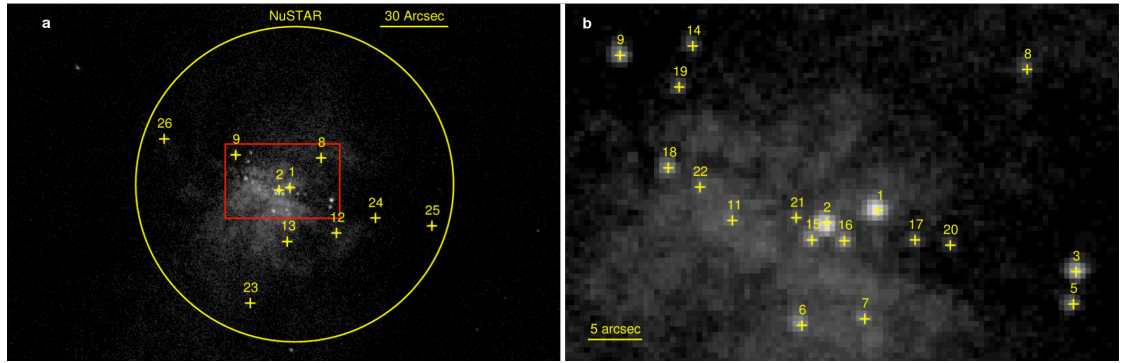

**Extended Data Figure 1 X-ray sources identified by Chandra in the central region of M82** Panel a shows a *Chandra* image of the central region of M82 from the observation taken coincident with *NuSTAR* ObsID 006. The green circle shows the 70″ radius region used to extract *NuSTAR* fluxes. Within this region, 24 discrete X-ray point sources are identified, including X-1 and X-2. The Panel b shows an expanded view of the crowded central region. Green crosses indicate the locations of identified point sources. We have used, where possible, the numbering from [13] (sources up to #15). After this we assign our own numerical identification (note that sources 4 and 10 from [13] are not detected in this observation).

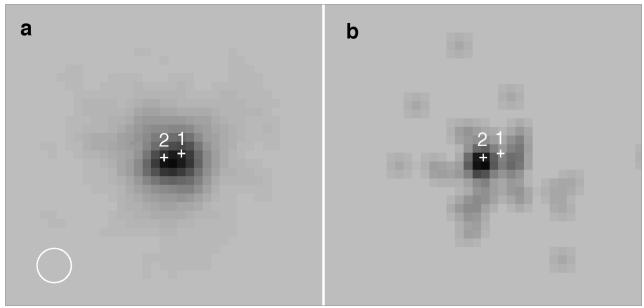

**Extended Data Figure 2 Swift Imaging of the region containing M82 X-1 and M82 X-2** Images of M82 X-1 and M82 X-2 seen by *Swift*. In grayscale are images obtained via the *Swift* automated processing over the 5 to 10 keV band for all of the observations during early February (2014/02/04 through 2014/02/11, panel a) and mid-March (2014/03/7 through 2014/03/11, panel b). The early February observations have 68.5 ks of exposure (mostly because of the increased cadence of observations due to the *Swift* monitoring of SN2014J), while the mid-March snapshot has 1.8 ks of exposure. The images are 1.5 arcminutes on a side and have been smoothed with a 2-pixel (4 arcsecond) Gaussian kernel (circle in lower left). The location of X-1 and X-2 are shown by the crosses in both panels. The late-time observation clearly shows a drop in the flux from X-1 and that the flux is dominated by X-2.